\documentclass[aps,prl,twocolumn,groupedaddress,floatfix,amsmath,showpacs]{revtex4-1} 	

\usepackage{graphicx}	
\usepackage{dcolumn}	
\usepackage{bm}		
\usepackage{amssymb}
\usepackage{natbib}
\usepackage{color}
\usepackage{soul} 	

\begin{document}

\title{Optical detection of spin transport in non-magnetic metals}

\author{F.~Fohr}
\author{S.~Kaltenborn}
\author{J.~Hamrle}
\altaffiliation{current address: Department of Physics, Ostrava University of Technology, 708 33 Ostrava, Czech Republic.}%
\author{H.~Schulthei\ss}
\altaffiliation{current address: Materials Science Division, Argonne National Laboratory, Argonne, Illinois 60439, USA.}
\author{A.~A.~Serga}
\author{H.~C.~Schneider}
\author{B.~Hillebrands}

\affiliation{%
Fachbereich Physik and Forschungszentrum OPTIMAS, Technische Universit\"at Kaiserslautern, D-67663 Kaiserslautern, Germany
}%

\author{Y.~Fukuma}%
\author{L.~Wang}%
\author{Y.~Otani}%

\affiliation{%
ASI, RIKEN, 2-1 Hirosawa, Wako 351-0198, and ISSP, University of Tokyo, 5-15-5 Kashiwanoha, Kashiwa 277-8581, Japan.
}%

\date{\today}

\begin{abstract}
We determine the dynamic magnetization induced in non-magnetic metal wedges composed of silver, copper and platinum by means of Brillouin light scattering (BLS) microscopy. The magnetization is transferred from a ferromagnetic Ni$_{80}$Fe$_{20}$ layer to the metal wedge via the spin pumping effect. The spin pumping efficiency can be controlled by adding an insulating interlayer between the magnetic and non-magnetic layer. By comparing the experimental results to a dynamical macroscopic spin-transport model we determine the transverse relaxation time of the pumped spin current which is much smaller than the longitudinal relaxation time.
\end{abstract}

\maketitle

\begin{figure}
\begin{center}
\includegraphics[width=0.48\textwidth]{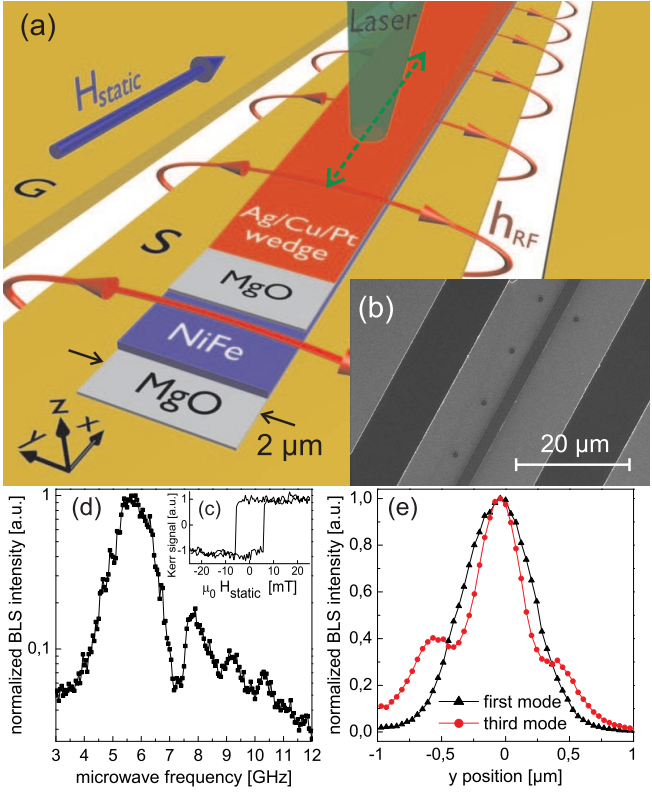}
\end{center}
\caption{(Color online) (a) Scheme of the sample layout and (b) SEM picture of the waveguide. A multilayer structure is prepared on top of a coplanar waveguide. A static magnetic field $\mathrm{\mu_0 H_\mathrm{static}}$ of 20\,mT is applied parallel to the signal line and perpendicular to the dynamic magnetic field $h_\mathrm{RF}$, which is caused by an alternating microwave current flowing through the coplanar waveguide. The magnetization in the Ni$_{80}$Fe$_{20}$ layer is excited by $h_\mathrm{RF}$ and spins are pumped into the metal wedge. (c) MOKE hysteresis loop to determine the saturation field in $x$-direction. (d) BLS spectrum taken on a pure Ni$_{80}$Fe$_{20}$ film at a static magnetic field of 20\,mT. (e) BLS scans across the structure for the first two maxima of (d) at microwave frequencies of 5.5\,GHz and 7.8\,GHz. The profiles correspond to the first and the third laterally standing spin wave mode. The second mode is not excited. }
\label{figure1} 
\end{figure} 

Spin current injection from a magnetic to a non-magnetic material is an important and central issue of magneto-electronics \cite{Prinz, Wolf}. There are several ways to realize such an injection. Spin currents can be generated by spin polarized charge currents \cite{MJohnson}, the spin Hall effect \cite{Hirsch}, or spin pumping \cite{Silsbee,Tserkovnyak}. The spin accumulation in the non-magnet can be detected indirectly by an increased damping in the injection layer \cite{Tserkovnyak, Gerrits} or it can be probed by the conversion of spin current into voltage in a lateral spin valve \cite{Jedemaa, Kimura1} or via the inverse spin Hall effect \cite{Saitoh, Kimura2}.

In this letter we demonstrate that the dynamic magnetization, which is induced via spin pumping into a non-magnetic material, can be observed directly by means of Brillouin light scattering (BLS) microscopy. We detect the spin polarization in metal wedges of Cu, Ag and Pt grown on top of a ferromagnetic layer (see Fig.~\ref{figure1}). Light from a laser is focussed on the surface of the wedge and the inelastically scattered light is collected as a function of the local wedge thickness. This light originates from the non-magnetic layer due to inelastic scattering from the spin polarization as well as from the magnetic layer below the wedge as long as the optical path length through the non-magnetic layer is smaller, or at least comparable to the optical absorption length. The magnetic layer is excited externally by the RF field of a coplanar waveguide (CPW) at the ferromagnetic resonance frequency, and generates the spin polarization in the wedge layer via the spin pumping effect. In the non-magnetic metals Cu, Ag and Pt,  the magnetic interaction between the spin-polarized free electrons is rather small, so that the macroscopic spin polarization, which is induced into the non-magnetic layer due to the spin pumping process, cannot be described in terms of collective eigen-excitations but reflects the magnetization dynamics in the ferromagnetic layer. The BLS signal arises from the forced oscillating macroscopic spin polarization so that frequency and wavevector are determined by  the magnons in the ferromagnetic film. Here we show that in the absence of a macroscopic magnetic ordering, interface impurities and roughness lead to a fast dephasing of the initially phase-aligned spins  in the non-magnetic layer. 


The CPW is prepared by means of maskless laser photolithography on an oxidized silicon substrate. It consists of a 300 nm gold layer with a signal line (S) of 20\,$\mu$m width separated from the ground planes (G) by a 10\,$\mu$m wide gap. A microwave current is applied to the CPW and generates an oscillating magnetic field in $y$-direction using a coordinate system as defined in Fig.~\ref{figure1}. To reach high microwave power in a wide frequency band and to prevent reflections, the microwave current is terminated by a load at the end of the CPW with impedance matching. 

On the CPW signal line a multilayer structure is deposited by electron beam evaporation. The multilayer has a width of 2\,$\mu$m, a length of 5\,mm and consists of:
($i$) a 7\,nm thick MgO layer that prevents the microwave current from flowing into the metal wedge, because this would create a complicated current distribution and an unpredictable magnetic field disturbing the CPW magnetic field;
($ii$) a 30\,nm thick Ni$_{80}$Fe$_{20}$ layer that is excited externally by the CPW dynamic magnetic field and serves as the pumping layer for the attached metal wedge;
($iii$) an optional  second 7\,nm thick MgO interlayer to block spin pumping from the Ni$_{80}$Fe$_{20}$ layer into the metal wedge;
($iv$) a metal wedge composed of either silver, copper or platinum. The optional MgO interlayer ($iii$) is used in a reference sample to separate the different contributions to the BLS intensity originating from the magnetic and the non-magnetic layer, respectively. In the sample without the MgO interlayer, spin pumping into the metal is expected to occur, whereas in the reference sample the pumping mechanism is blocked by the MgO layer. The latter is insulating but optically transparent, and therefore does not affect the detection by the probing laser light.

For absolute height calibration of the metal wedge, the scan position in $x$-direction is calculated into a total thickness. The topography was scanned in $y$-direction with a mechanical profilometer for different points along the wedge, and for each of these profiles the thickness of the multilayer was extracted with the CPW level as reference level.

Figure \ref{figure1}(d) shows the BLS spectrum taken on a pure Ni$_{80}$Fe$_{20}$ film at an applied field of 20\,mT in $x$-direction. The first and most pronounced maximum is visible at a frequency of 5.5\,GHz but several other maxima develop at higher frequencies corresponding to higher laterally standing spin wave modes across the stripe \cite{Demokritov, Bayer}. 

A spatially uniform precession cannot be excited in a 2\,$\mu$m wide stripe due to pinning effects at the boundaries. Standing spin waves build up across the width of the stripe in $y$-direction and the spin pumping efficiency becomes dependent on this coordinate (Fig.~\ref{figure1}(e)). The dynamic magnetization in the non-magnetic layer experiences additional dephasing due to the mixing of components pumped with different initial phases from neighbouring antinodes of higher-order standing waves. To minimize this contribution to the dephasing, only the first standing spin wave mode, excited at a microwave frequency of 5.5\,GHz and an external field of 20\,mT was used in our BLS measurements.  

\begin{figure}
\begin{center}
\includegraphics[width=0.4\textwidth]{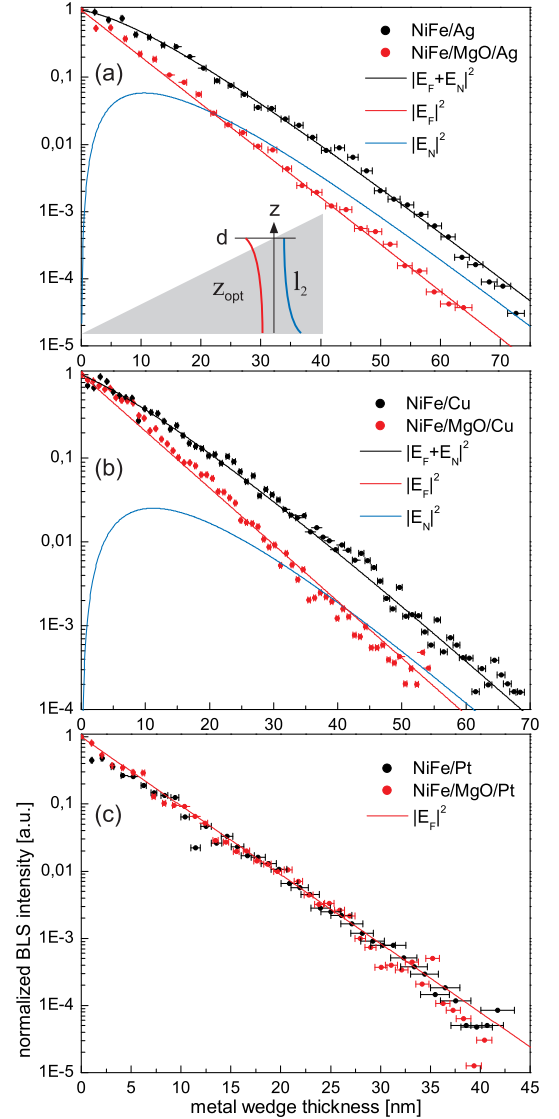}
\end{center}
\caption{(Color online) BLS scan data of the silver (a), the copper (b) and the platinum (c) wedged sample. Each graph shows the measurement of the main sample with active spin pumping (black dots) and the respective reference sample with blocked spin pumping (red dots). A difference between main and reference sample is only visible for silver and copper but not for platinum (see text). The error bars reflect the uncertainty in thickness determination by the mechanical profilometer. The black and red lines are fits of the scan data according to Eq.~(\ref{equationa}). The evolution of the pure spin part of the BLS signal (blue line) is derived from the fitting parameters. The inset in (a) shows schematically that the optical absorption length $\mathrm{z_\mathrm{opt}}$ as well as the decay length of the transverse component $l_{2}$ contribute to the total BLS intensity probed by the laser.}
\label{figure2} 
\end{figure}

In Fig.~\ref{figure2} the measured BLS intensities of the maximum of the first mode are shown for different scan positions in $x$-direction. With increasing wedge thickness the BLS signal decays exponentially over a range of almost four decades in intensity. In the silver (Fig.~\ref{figure2}(a)) and the copper wedge (Fig.~\ref{figure2}(b)) the slopes of the exponential decay are different for the main (black dots) and the reference sample (red dots) whereas in the platinum sample both slopes are the same within the error bars (Fig.~\ref{figure2}(c)). 

The origin of the difference in silver and copper is the additional contribution to the BLS intensity due to the spin polarization pumped from the underlying Ni$_{80}$Fe$_{20}$ layer. While the BLS signal of the main sample is determined by the optical decay from the signal originating in Ni$_{80}$Fe$_{20}$ as well as by the induced magnetization in copper, the signal from the sample with the MgO interlayer, which prevents spin pumping, contains only the signal originating from the Ni$_{80}$Fe$_{20}$ layer. In platinum this effect is not observable because the injected spin angular momentum is immediately transferred from the spin system to the lattice due to the high spin orbit interaction. 


The total BLS intensity depends on the thickness of the metal wedge and consists of two contributions: One is due to the precessing magnetization in the ferromagnet, the other originates from the pumped spin polarization in the metal. The total BLS intensity is proportional to:
\begin{equation}
|E_\mathrm{F}+E_\mathrm{N}|^2=|E_\mathrm{F}|^2+2\operatorname{Re}\left(E_\mathrm{F}E^*_\mathrm{N}\right)+|E_\mathrm{N}|^2
\label{equationa}
\end{equation}
where $E_\mathrm{F}$ and $E_\mathrm{N}$ are the electric fields of the probe laser light scattered inelastically in the magnetic and the non-magnetic layer, respectively. Inside the metal wedge of thickness $d$, i.e. for $0<z<d$, the profile of the electric field originating from the incident light can be expressed as a damped wave (see inset of Fig.~\ref{figure2}). 
\begin{equation}
E_\mathrm{F}=E_\mathrm{F,0}  \exp{\left(2i\tilde{n}kd\right)}
\label{equationb}
\end{equation}

Here $\tilde{n}=n+{\it i}\kappa$ is the complex refractive index and $k=\omega/c$ is the vacuum wavevector of light. The factor 2 in the exponent takes into account that the BLS setup is prepared in backscattering geometry, and thus the light is passing through the structure twice. 

The amplitude of the backscattered light from the non-magnetic layer is a sum of contributions originating from different depths of the wedge, weighted by the decaying probe light amplitude as well as by the decaying contribution of the spin polarization to the scattered light:
\begin{equation}
E_\mathrm{N}=\int \limits_{0}^{d}  E_\mathrm{N,0}       {\exp{\left[2i\tilde{n}k\left(d-z\right)\right]}    \exp{\left(-\frac {z}{l_{2}}\right)} dz }
\label{equationc}
\end{equation}
Here $l_{2}$ is the characteristic decay length of the dynamic spin polarization that gives rise to the BLS signal. 

In Fig.~\ref{figure2} the fit curves (black and red lines) and the extracted BLS intensity for the pure spin part (blue line), obtained by using Eqs.~(\ref{equationa})-(\ref{equationc}), are shown in addition to the measurement data. The fitting parameters in the simulation are the ratio of the field strengths $E_\mathrm{N,0}$ and $E_\mathrm{F,0}$ at the Ni$_{80}$Fe$_{20}$/Cu interface, the complex refractive index $\tilde{n}$ in the metal and the decay length $l_{2}$. The results of this simulation as well as the parameters of the fits in Fig.~\ref{figure2} are summarized in Tab.~\ref{tab:fit}. The contribution of the induced magnetization $|E_\mathrm{N}|^2$ to the BLS signal is at maximum for wedge thicknesses below 10\,nm, even if the BLS signal originates from the Ni$_{80}$Fe$_{20}$ layer, i.e. $|E_\mathrm{F}|^2$ is dominant. Above wedge thicknesses of 20\,nm in silver (45\,nm in copper) $|E_\mathrm{N}|^2$ becomes dominating over $|E_\mathrm{F}|^2$. 

With knowledge of $l_{2}$, a characteristic relaxation time $T_{2}$ can be calculated using wave-diffusion equations for the macroscopic spin density $\vec{\rho}_\mathrm{s}(z,t)=\vec{\rho}_{\uparrow}(z,t)-\vec{\rho}_{\downarrow}(z,t)$ and the spin-current density $\vec{J}_\mathrm{s}(z,t)=\vec{J}_{\uparrow}(z,t)-\vec{J}_{\downarrow}(z,t)$: 
\begin{equation}
\frac{\partial\vec{\rho}_\mathrm{s}(z,t)}{\partial t}=-\gamma\vec{\rho}_\mathrm{s}(z,t)\times\vec{B}-\frac{\partial\vec{J}_\mathrm{s}(z,t)}{\partial z}-\left.\frac{\partial\vec{\rho}_\mathrm{s}(z,t)}{\partial t}\right|_{\mathrm{int}},\label{eq:continuity noncol}
\end{equation}
and
\begin{equation}
\vec{J}_\mathrm{s}(z,t)=-D\frac{\partial\vec{\rho}_\mathrm{s}(z,t)}{\partial z}-\tau_\mathrm{e}\gamma\vec{J}_\mathrm{s}(z,t)\times\vec{B}-\tau_\mathrm{e}\frac{\partial\vec{J}_\mathrm{s}(z,t)}{\partial t}.\label{eq:diffusion noncol}
\end{equation} Here,  $\tau_\mathrm{e}$ denotes the momentum relaxation time, $\gamma$ is the absolute value of the electron $\left(g\approx2\right)$ gyromagnetic ratio, $\vec{B}$ is the magnetic field and $D$ is the diffusion constant.

As a generalization of Ref.~\cite{Zhu}, we include different longitudinal (or spin-lattice) relaxation times $T_{1}$ and transverse (or spin-spin) relaxation times $T_{2}$ in the interaction contribution in Eq.~(\ref{eq:continuity noncol}): $\left.\left(\partial \rho_\mathrm{s}\left(x\right)/\partial t\right)\right|_{\mathrm{int}}=\rho_\mathrm{s}\left(x\right)/T_{1}$ and $\left.\left(\partial \rho_\mathrm{s}\left(y,z\right)/\partial t\right)\right|_{\mathrm{int}}=\rho_\mathrm{s}\left(y,z\right)/T_{2}$. 

The dynamical components of $\vec{\rho}_\mathrm{s}(z,t)$ in Eq.~(\ref{eq:continuity noncol}) decay with $T_{2}$ and the static component decays with $T_{1}$. The latter is not accessible to Brillouin light scattering and is therefore neglected in the following.

It can be shown  \cite{Zhu} that the relaxation time $T_{2}$ depends on the corresponding decay length $l_{2}$ via:
\begin{equation}
l_{2}=\frac{v_{\mathrm{F}}}{\sqrt{3}}\sqrt{\tau_{\mathrm{e}} T_{2}}
\label{lengths}
\end{equation}
where $v_{\mathrm{F}}$ is the Fermi velocity. 

To obtain the transverse relaxation time, we solve Eqs. (\ref{eq:continuity noncol}) and (\ref{eq:diffusion noncol}) numerically by using $T_{2}$ as a fit parameter to match the experimentally determined value for $l_{2}$ (see Tab.~\ref{tab:fit}). According to Ref.~\cite{Ashcroft} we use $\tau_{\mathrm{e}}\left(\mathrm{Cu}\right)=25$\,fs and $\tau_{\mathrm{e}}\left(\mathrm{Ag}\right)=40$\,fs for the momentum relaxation time and $v_{\mathrm{F}}\left(\mathrm{Cu}\right)=1.57$\,nm/fs and $v_{\mathrm{F}}\left(\mathrm{Ag}\right)=1.39$\,nm/fs for the Fermi velocity at room temperature. The influence of the small external magnetic field of 20\,mT and the injection frequency of 5.5\,GHz on the decay length is negligible in the calculations. This result is also confirmed by our BLS measurements of the copper sample: The decay is unchanged within the error bars at an injection frequency of 9.3 GHz and an applied field of 70\,mT.  

\begin{table} 
\caption{\label{tab:fit} Fitting parameters of the BLS data in Fig.~\ref{figure2} and the resulting transverse relaxation time $T_2$ according to the macroscopic spin wave-equation.}
\begin{ruledtabular}
\begin{tabular}{llllll}
Normal \\ metal & $\tilde{n}$ & $\tilde{n}_\mathrm{lit}$ in \cite{Palik} & $l_2$ (nm)  &  $E_\mathrm{F,0}/E_\mathrm{N,0}$ & $T_2$ (fs)  \\
 \hline
Ag	 &  0.13+i3.4 	& 0.13+i3.2 	& 9 $\pm{1}$  	& 16		&  3 $\pm{1}$ \\
Cu 	& 1.07+i3.3  	& 1.07+i2.6  	& 10 $\pm{1}$  	& 26 		& 5 $\pm{1}$ \\
Pt  	&  2.08+i5.2  	& 2.08+i3.6  	& 0 $\pm{2}$   	&  $>20$ 	& 0 $\pm{1}$ \\
\end{tabular}
\end{ruledtabular}
\end{table}

Note that the accepted value of the longitudinal relaxation time $T_{1}$, which is of the order of a few picoseconds \cite{Bass, Jedema}, exceeds the value of  $T_{2}$ determined by our BLS measurements by three orders of magnitude. This is a remarkable result because $T_{2}$ is usually considered equivalent to $T_{1}$ \cite{Pines}. The magnitude of the discrepancy between $T_{1}$ and $T_{2}$ suggests an extrinsic effect acting differently on the transverse and the longitudinal component of the induced magnetization. 

In order to understand this discrepancy we propose a relaxation mechanism, which is  based on absorption of the  transverse spin current  due to magnetic impurities at the interface and dephasing of the transverse spin current  in inhomogeneous magnetostatic fields arising from the interface roughness. Indeed, magnetic impurities at the interface were detected by means of secondary  ion mass spectroscopy (SIMS). The magnetization of the paramagnetic impurities is aligned along the direction of the external magnetic field, so that the transverse spin current is effectively absorbed by exerting a maximum torque on the magnetization of the impurities. In case of  spin current transmission from a nonmagnet into a ferromagnet \cite{Stiles}, this effect is even more pronounced due to the influence of three processes:  ($i$) spin-dependent reflection and transmission, ($ii$) rotation of reflected and transmitted spins, and ($iii$) spatial precession of spins in the ferromagnet. In our case, the spin current is transmitted from the ferromagnetic to the non-magnetic layer, so that the interface effect ($ii$) as well as effect ($iii$), which is based on the different kinetic energies of spin-up and spin-down electrons due to exchange splitting, can be neglected in our considerations. Therefore, only effect ($i$)  contributes to relaxation and can be understood by expressing the transverse spin current as a linear combination of spin-up and spin-down components, which have different reflection and transmission amplitudes when they scatter on the paramagnetic impurities. As a result, the transverse spin current has a non-zero propagation length in contrast to the immediately absorbed spin current in  \cite{Stiles}.

The analysis of the Ni$_{80}$Fe$_{20}$ layer topography via atomic force microscopy (AFM) reveals that Ni$_{80}$Fe$_{20}$ clusters with a grain size of several tens of nanometers and an average Ni$_{80}$Fe$_{20}$ surface roughness of 1.9\,nm. This surface roughness is transferred to the Ni$_{80}$Fe$_{20}$/nonmagnet interface and causes spatially inhomogeneous magnetic fields, so that the spins, which precess  initially  with the same frequency and phase, experience additional dephasing due to different precession frequencies in these local magnetic fields \cite{Dash}. 

In conclusion we have determined for the first time the existence of magnetization in non-magnetic metals by optical means. The transverse spin current is directly accessible to Brillouin light scattering microscopy and decays faster than the longitudinal spin current due to absorption and dephasing at the interface. 

\begin{acknowledgments}
We acknowledge the financial support by the Deutsche Forschungsgemeinschaft and the Japan Science and Technology Agency. We also thank W. Bock from the IFOS institute for the SIMS measurements.
\end{acknowledgments}

\end{document}